\documentclass[conference, letterpaper]{IEEEtran}
\pdfoutput=1
\usepackage[pass]{geometry}
\usepackage{cite}

\ifCLASSINFOpdf
   \usepackage[pdftex]{graphicx}
   \graphicspath{{../pdf/}{../jpeg/}}
   \DeclareGraphicsExtensions{.pdf,.jpeg,.png}
\else
   \usepackage[dvips]{graphicx}
   \graphicspath{{../eps/}}
   \DeclareGraphicsExtensions{.eps}
\fi
\usepackage{tikz}
\usepackage{textcomp}
\usepackage{hyperref}
\usepackage{lipsum}

\newcommand\copyrighttext{
  \footnotesize Copyright \textcopyright 2015  IEICE. \textit{Antennas and Propagation (ISAP), International Conference on}, Nov. 2015.}
\newcommand\copyrightnotice{%
\begin{tikzpicture}[remember picture,overlay]
\node[anchor=south,yshift=10pt] at (current page.south) {\fbox{\parbox{\dimexpr\textwidth-\fboxsep-\fboxrule\relax}{\copyrighttext}}};
\end{tikzpicture}%
}

\usepackage[cmex10]{amsmath}
\usepackage{url}

\usepackage{booktabs}

\begin{document}
%

\title{Measured Aperture-Array Noise Temperature of the Mark II Phased Array Feed for {ASKAP}}

\author{\IEEEauthorblockN{A.~P. Chippendale, A.~J. Brown, R.~J. Beresford, G.~A. Hampson,
 R.~D. Shaw, D.~B. Hayman,\\ A. Macleod, A.~R. Forsyth, S.~G. Hay,
M.~Leach, C.~Cantrall, M.~L. Brothers and A.~W. Hotan}
\IEEEauthorblockA{CSIRO Astronomy and Space Science\\
PO Box 76, Epping, NSW 1710, Australia\\
Email: Aaron.Chippendale@csiro.au}}


%


\maketitle
\copyrightnotice

\begin{abstract}
We have measured the aperture-array noise temperature of the first Mk.~II phased array feed that CSIRO has built for the Australian Square Kilometre Array Pathfinder telescope.  As an aperture array, the Mk.~II phased array feed achieves a beam equivalent noise temperature less than 40~K from 0.78~GHz to 1.7~GHz and less than 50~K from 0.7~GHz to 1.8~GHz for a boresight beam directed at the zenith.  We believe these are the lowest reported noise temperatures over these frequency ranges for ambient-temperature phased arrays.  The measured noise temperature includes receiver electronics noise, ohmic losses in the array, and stray radiation from sidelobes illuminating the sky and ground away from the desired field of view. This phased array feed was designed for the Australian Square Kilometre Array Pathfinder to demonstrate fast astronomical surveys with a wide field of view for the Square Kilometre Array. 

\end{abstract}


%
\IEEEpeerreviewmaketitle

\section{Introduction}
CSIRO developed the Mk.~II phased array feed (PAF) \cite{Hampson2012} for the Australian Square Kilometre Array Pathfinder (ASKAP) \cite{DeBoer2009} to demonstrate fast astronomical surveys with a wide field of view \cite{Johnston2008} for the Square Kilometre Array (SKA).  The SKA is an international project to build the world's largest radio telescope with a square kilometre of collecting area \cite{Dewdney2009}.

Over the next two years, CSIRO will install Mk.~II PAFs on thirty 12~m parabolic reflector antennas of the ASKAP telescope.  ASKAP uses digital beamforming to simultaneously process 36 beams from each PAF, increasing the field of view of each reflector antenna by a factor of 30 over the same antenna with a single-pixel feed \cite{Bunton2010, Hay2013}.

\begin{figure}
\centering
\includegraphics[width=0.8\columnwidth]{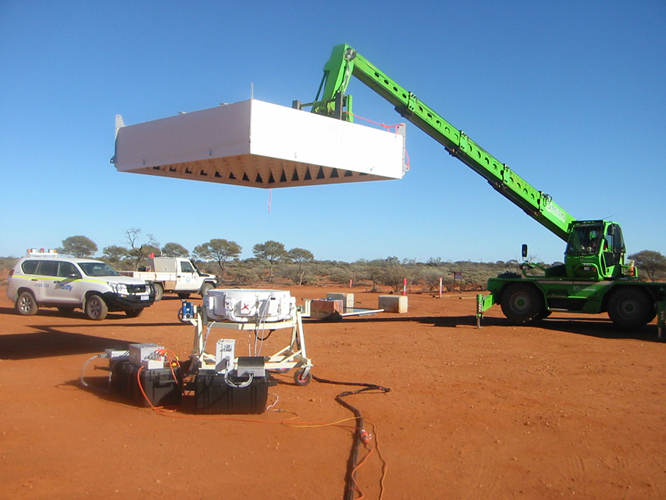}
\caption{Aperture-array Y-factor measurement of the Mk.~II ASKAP PAF near antenna 29 at the MRO.  The array is supported on the ground so that it points face-up at the zenith. A telehandler is used to position a large microwave absorber load over and alternately away from the array. }
\label{fig:meas}
\end{figure}

This paper presents aperture-array noise temperature measurements of the first Mk.~II PAF.  We made these measurements at the Murchison Radio-astronomy Observatory (MRO), as shown in Fig. \ref{fig:meas}, using an accurate Y-factor technique developed to verify the noise performance of prototype phased array antennas for ASKAP and the SKA \cite{Chippendale2014, Hayman2014}.  The Y-factor is the ratio of beamformed power between observations of two scenes with different but known brightness temperatures.   

\begin{figure}
\centering
\includegraphics[width=0.99\columnwidth]{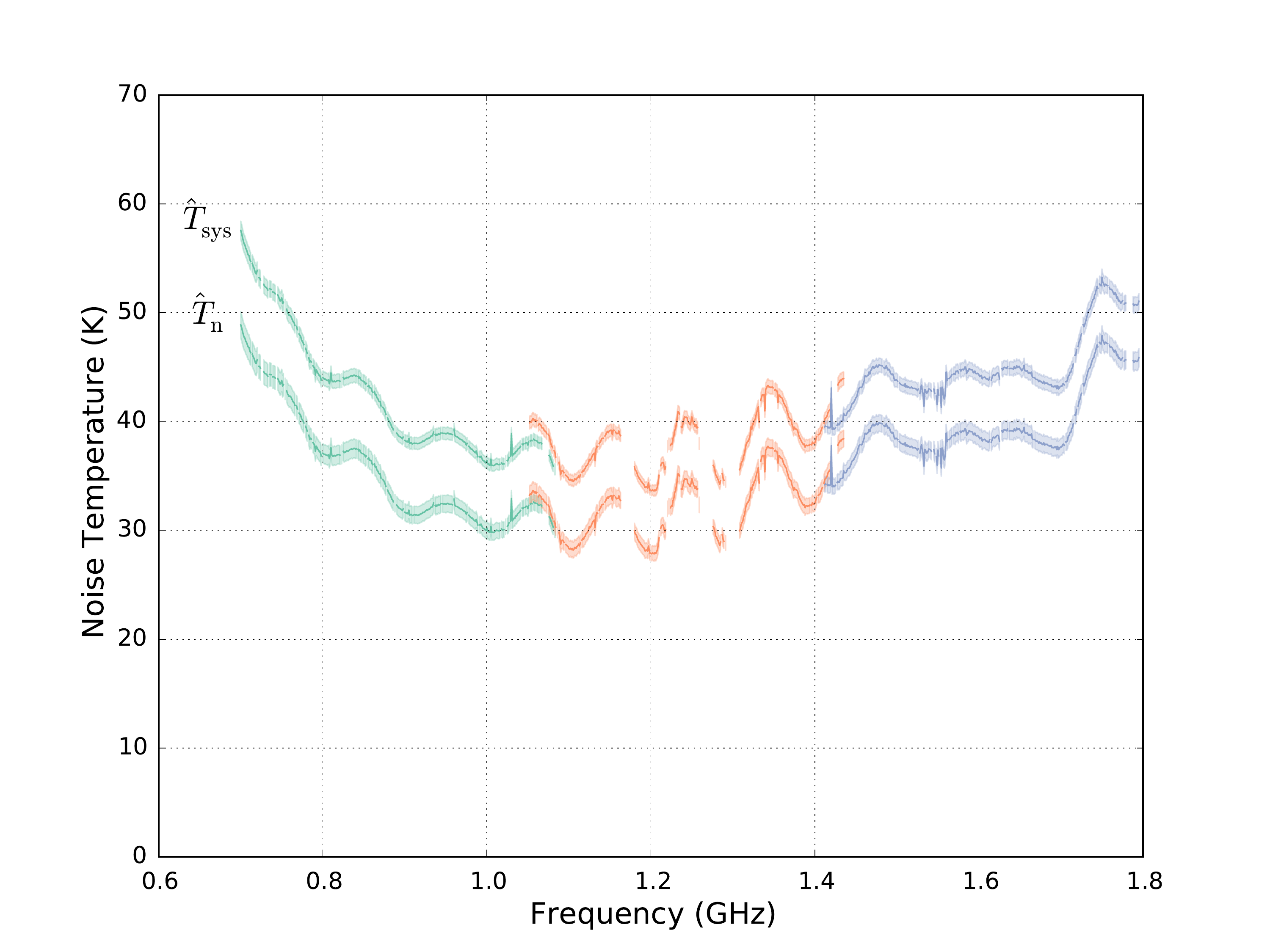}
\caption{Measured beam equivalent noise temperatures of the Mk.~II ASKAP PAF for a boresight beam with maximum $S/N$ weights.  The upper curve is the system temperature $\hat{T}_\text{sys}$ with all noise contributions.  The lower curve is the partial noise temperature $\hat{T}_\text{n}$ that does not include the background radio sky and atmospheric contributions via the main beam.  The shaded band around each line represents the measurement uncertainty calculated following \cite{Chippendale2014}.  The three colours represent separate measurements in different sampling bands.}
\label{fig:result}
\end{figure}

Fig. \ref{fig:result} shows that the Mk.~II PAF achieves a beam equivalent noise temperature less than 40~K from 0.78~GHz to 1.7~GHz and less than 50~K from 0.7~GHz to 1.8~GHz for a boresight beam directed at the zenith.  We believe these are the lowest reported noise temperatures over these frequency ranges for ambient-temperature phased arrays.  Our results are comparable to the Low-Noise Tile developed by ASTRON \cite{Woestenburg2014} to demonstrate improved sensitivity for a Mid Frequency Aperture Array element of the SKA and the thick Vivaldi element array developed by NRC \cite{Veidt2015} for use as a PAF on a reflector antenna.   

Sensitivity of the Mk.~II PAF has been significantly improved over the Mk. I which had a steep degradation above 1.2 GHz, becoming less than half as sensitive above 1.4~GHz \cite{Schinckel2011}. The improvement was achieved by enhancing antenna array and low-noise amplifier (LNA) designs following very careful measurements of LNA signal and noise parameters \cite{Shaw2012}. 

Both Mk. I and Mk.~II ASKAP PAFs are based on a connected-element ``chequerboard'' array \cite{Hay2008} that is dual-polarized, low-profile, and inherently wide-band.  The Mk.~II PAF tested here has since been installed at the focus of an ASKAP antenna where it has demonstrated good on-dish sensitivity from 700~MHz to 1.8~GHz \cite{Chippendale2015}.  Work is also underway to develop LNAs with newer transistors and lower minimum noise temperatures \cite{Shaw2015}.  These could be used to build more sensitive PAFs for the SKA or to upgrade ASKAP.

\section{Aperture-Array Measurement System}
We followed the Y-factor noise measurement technique described in \cite{Chippendale2014}, but with second generation ASKAP digital hardware on a new test-site at the MRO in Western Australia.  Fig. \ref{fig:meas} shows the array under test on the ground near ASKAP antenna 29, which is located at 26.6902147896~S, 116.6371346089~E \cite{CSIRO2013}.  Antenna 29 is cabled with an additional fibre junction box so that we may test an array on the ground as an aperture array (this paper) or at the focus of the reflector \cite{Chippendale2015}.  This system is supported by a full single-antenna ASKAP digital receiver \cite{Brown2014} and beamformer \cite{Hampson2014} in the MRO control building.  

The 188 radio-frequency (RF) signals from the ``chequerboard'' array are directly modulated onto optical fibre links that traverse 1.4~km to the digital receiver in the control building.  The digital receiver \cite{Brown2014} directly samples 192 signals (188 from the PAF and 4 spare) and then uses an oversampled poly-phase filter bank \cite{Tuthill2012} to divide these signals into 1~MHz channels.  Each directly-sampled band has approximately 600~MHz RF bandwidth, but only 384~MHz are passed to the beamformer in the current configuration.  A particular contiguous 384~MHz sub-band is selected for beamforming via a software command to set the centre frequency.  

Table \ref{tab:measparams} lists the frequency ranges of the three sampling bands used to cover the ASKAP band and the particular 384~MHz sub-bands selected for the measurements in this paper.  A fourth sampling band covering 600~MHz to 700~MHz is available for experimentation with arrays that work at lower frequencies as planned for the SKA.

For each measurement, a box lined with microwave absorber was positioned over the array under test using a telehandler as shown in Fig. \ref{fig:meas}.  The telehandler arm was then swung away so that the array could observe the unobstructed sky.  The 3~m~$\times$~3~m absorber box is made with conductive carbon fibre.  It is 700~mm tall and is lined with 600~mm long pyramidal cones of microwave absorber ($\text{Franko}_\text{Sorb}\textsuperscript{\textregistered}\text{P}600$) with 300~mm~$\times$~300~mm square bases installed tips-down.  The manufacturer quotes reflectivity of -30~dB at 900~MHz for normal incidence.    

The ``chequerboard'' array surface was 1,310~mm above the ground. The vertical distance between the absorber cone tips and array surface was between 500~mm and 900~mm (cf. 1,270~mm in \cite{Chippendale2014}).  Compared to the 2.44~m~$\times$~2.90~m wheel-on-track absorber at Parkes \cite{Chippendale2014}, it was not as easy to control the height and lateral position of the absorber over the array with the telehandler at the MRO.  Although with further practice we could probably position the absorber with 100~mm accuracy.

The uncertain absorber height should have a small effect on the measurements presented here because we are using a larger absorber much closer to a larger and more directive array than in \cite{Chippendale2014}.  This means the absorber illumination efficiency $\alpha$ will be higher so relative errors in $\alpha$ due to absorber height variation, and their impact on noise temperature measurements, will be smaller.  We also limited measurements to times when the Galactic centre was below the horizon. 

\begin{figure}
\centering
\includegraphics[width=0.8\columnwidth]{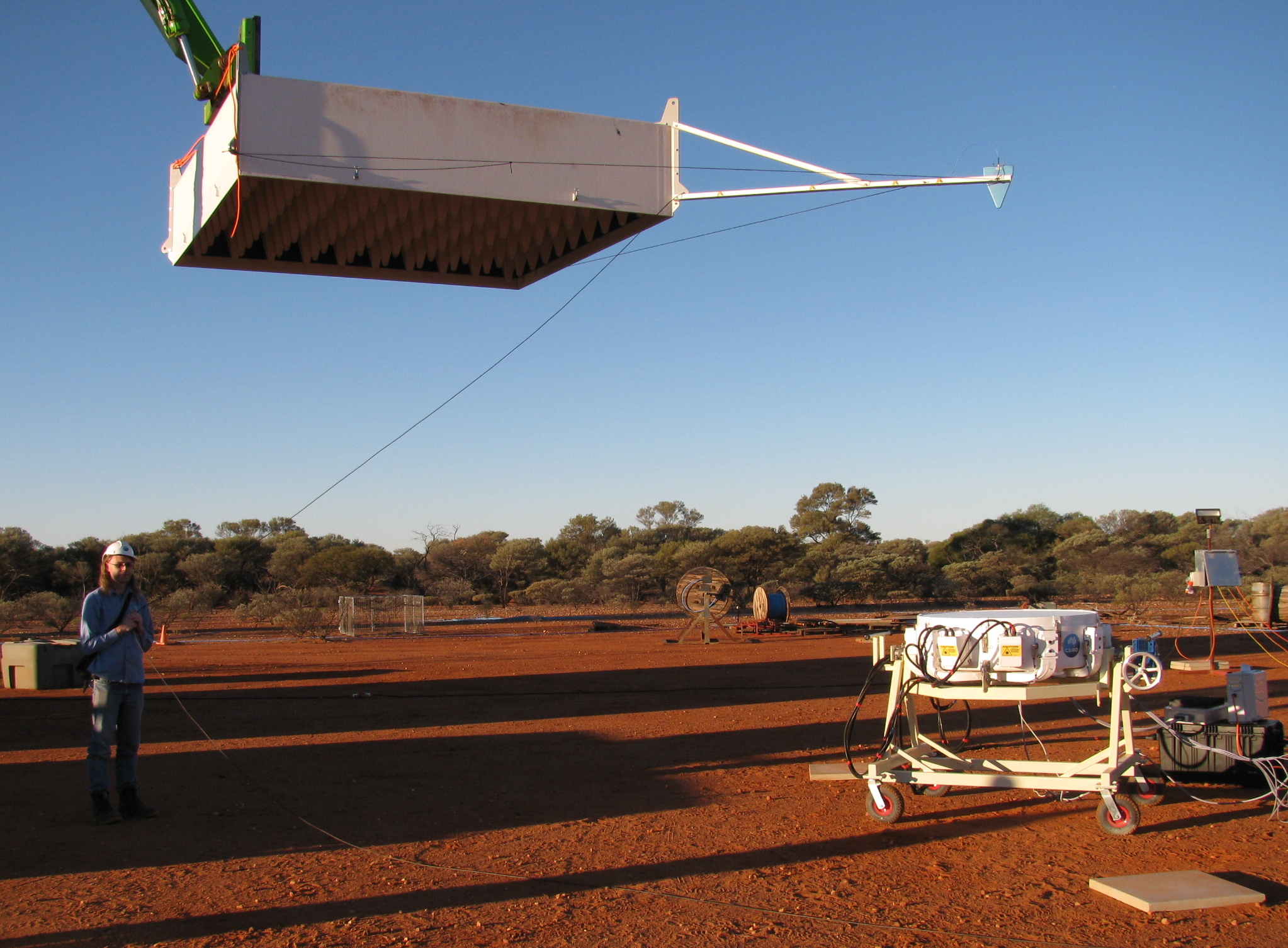}
\caption{Broadband noise is radiated into the array under test from an antenna suspended directly above the centre of the array.  Correlating a copy of this noise with the received signal from each array port allows the calculation of weights to steer a beam at the zenith.}
\label{fig:radiator}
\end{figure}

The direction and polarisation of the beam were controlled via measurements of broadband noise radiated from a reference antenna located directly above the array under test as in \cite{Chippendale2014} and \cite{Hayman2014}.  At the MRO we mounted a log-periodic dipole array antenna (Aaronia HyperLOG 7025) on a plastic beam extending 2.4~m from one corner of the absorber box as shown in Fig. \ref{fig:radiator}.  We used the telehandler to place the noise-reference antenna over the boresight of the array, but at a greater height of 2.7~m above the array so that the reference signal would have a flatter wave-front at the array's surface.

\section{Aperture-Array Noise Measurements}
The primary measurement was the Y-factor ratio between the beamformed power $P_\text{hot}$ received from the the microwave absorber at ambient temperature (nominally 294~K) and $P_\text{cold}$ received from the unobstructed sky (nominally 8~K)
\begin{equation}
  Y = \frac{P_{\text{hot}}}{P_{\text{cold}}} = \frac{\mathbf{w}^H\mathbf{R}_{\text{hot}}\mathbf{w}}{\mathbf{w}^H\mathbf{R}_{\text{cold}}\mathbf{w}}.
  \label{eq:yfact}
\end{equation}
Here $\mathbf{w}$ is a vector of beamformer weights and $\mathbf{R}_\text{hot}$ and $\mathbf{R}_\text{cold}$ are the sample array covariance matrices measured towards the absorber and unobstructed sky respectively.  The RHS of Equation \eqref{eq:yfact} is evaluated offline using covariance matrices calculated by the ASKAP beamformer \cite{Hampson2014} according to 
\begin{equation}
  \mathbf{R} = \frac{1}{L}\sum_{n=1}^{L}\mathbf{x}(n)\mathbf{x}^{H}(n)
  \label{eq:scm}
\end{equation}
where $\mathbf{x}(n)$ is the $n^{\text{th}}$ time sample of the column vector of 188 complex array-port voltages $\mathbf{x}(t)$.   Covariance matrices, beamformer weights, and beamformed Y-factors are calculated independently for each 1~MHz digital receiver channel.   We used maximum signal-to-noise ratio ($S/N$) weights for a beam directed at the zenith \cite{Lo1966, Widrow1967, Chippendale2014} 
\begin{equation}
  \mathbf{w} =\mathbf{R}_\text{cold}^{-1}\mathbf{r}_{xd}.
  \label{eq:maxsnrweights}
\end{equation}
where $\mathbf{r}_{xd} $ is the sample cross-correlation vector
\begin{equation}
  \mathbf{r}_{xd} = \frac{1}{L}\sum_{n=1}^{L}\mathbf{x}(n)d^{*}(n)
\end{equation}
and $d(n)$ is a sampled copy of the broadband noise voltage radiated into the array to define the look direction and polarisation state of the beam \cite{Chippendale2014}.

We define sky-referenced beam equivalent system noise temperature \cite{Chippendale2014, Warnick2010} 
\begin{equation}
\label{eq:tsys}
\hat{T}_\text{sys} = T_{\text{ext},\text{sky}} + T_{\text{ext},\text{gnd}} + (T_{\text{loss}} + T_{\text{rec}})/\eta_\text{rad} 
\end{equation} 
where $T_{\text{ext},\text{sky}}$ is the component of noise temperature due to sky emission including the background radio sky and Earth's atmosphere, $T_{\text{ext},\text{gnd}}$ is due to ground emission, $T_{\text{loss}}$ is due to ohmic losses in the array, $T_{\text{rec}}$ is due to receiver electronics noise, and $\eta_\text{rad}$ is the beam radiation efficiency.  We calculate $\hat{T}_\text{sys}$ from the beamformed Y-factor via \cite{Chippendale2014}
\begin{equation}\label{eq:tsys-via-y}
\hat{T}_{\text{sys}} = \frac{T_{\text{sys}}}{\eta_\text{rad}} = \frac{\alpha T_{\text{abs}} - T_{\text{ext},\text{sky}(A)}}{Y-1}
\end{equation}
where $\alpha$ is the absorber illumination efficiency describing how well the absorber load fills the beam, $T_\text{abs}$ is the physical temperature of the absorber, and $T_{\text{ext},\text{sky}(A)}$ is the noise temperature component due to emission from the region of sky $A$ blocked by the absorber when viewed from the array under test.  

As a step towards isolating $T_\text{loss}$ and $T_\text{rec}$ to ease comparison of receivers measured at different sites and times, we also define the partial beam equivalent noise temperature \cite{Chippendale2014}
\begin{equation}
\label{eq:tpart}
\hat{T}_\text{n} = T_{\text{ext},\text{sky}(B)} + T_{\text{ext},\text{gnd}} + (T_\text{loss} + T_\text{rec})/\eta_\text{rad}.
\end{equation}
This is essentially $\hat{T}_\text{sys}$ less the external sky-noise $T_{\text{ext},\text{sky}(A)}$ from the region of sky blocked by the absorber, the remaining sky-noise $T_{\text{ext},\text{sky}(B)}$ is from the region $B$ that is not blocked by the absorber.  The partial noise temperature is calculated from the beamformed Y-factor via \cite{Chippendale2014}
\begin{equation}
\begin{split}
  \hat{T}_{\text{n}} = \frac{T_{\text{n}}}{\eta_{\text{rad}}} &=  \frac{\alpha T_\text{abs}-YT_{\text{ext},\text{sky}(A)}}{Y-1}.
  \end{split}
  \label{eq:noisetemp}
\end{equation}

\begin{figure}
\centering
\includegraphics[trim=60 10 70 55 ,clip,width=0.80\columnwidth]{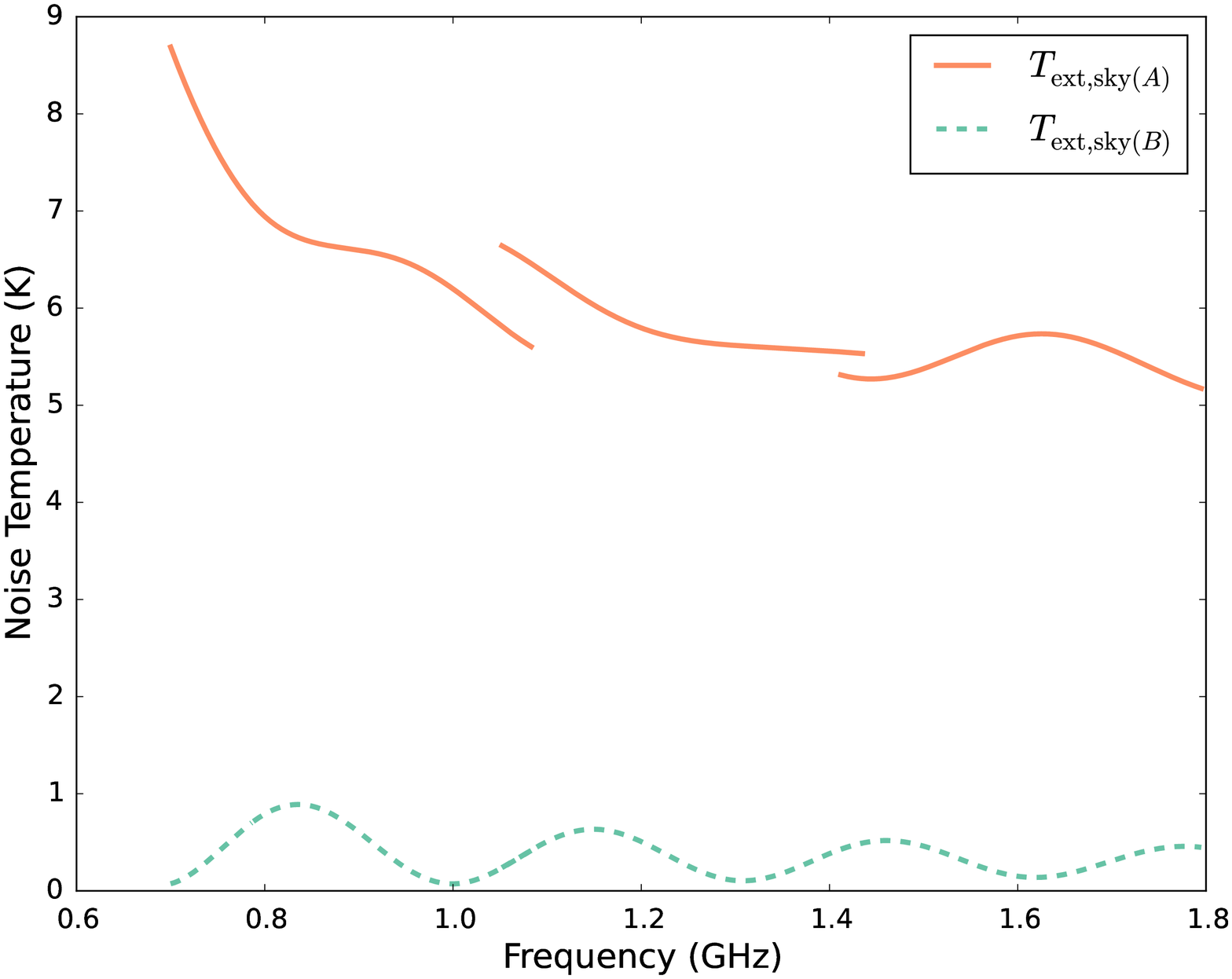}
\caption{Noise temperature contributions from the sky estimated via \cite{Chippendale2014} including the background radio sky, the Sun, and the Earth's atmosphere.  $T_{\text{ext},\text{sky}(A)}$ represents emission from sky blocked by the absorber and $T_{\text{ext},\text{sky}(B)}$ is the stray component seen in both absorber and open-sky measurements.  These were calculated for the absorber 500~mm above the array and only vary by 1~K if the absorber is 900~mm above the array.  Discontinuities arise because the three sub-bands were measured at different times (see Table \ref{tab:measparams}) and the array sees different parts of the Galaxy's emission as the Earth rotates.   }
\label{fig:tsky}
\end{figure}

The upper and lower curves in Fig. \ref{fig:result} respectively show the resulting $\hat{T}_\text{sys}$ and $\hat{T}_\text{n}$ for the Mk.~II ASKAP PAF using $\alpha=0.96\pm0.04$, $T_{\text{ext},\text{sky}}$ as estimated in Fig. \ref{fig:tsky}, and the infrared thermometer measurements of $T_\text{abs}$ in Table \ref{tab:measparams}. The value of $\alpha$ was calculated according to \cite{Chippendale2014} assuming uniform weights and isotropic element patterns to estimate the array beam.  It is difficult to accurately calculate $T_{\text{ext},\text{sky}}$ due to poor knowledge of the beam.  However, provided the Galactic centre and Sun don't enter near sidelobes, the result should be close to the estimates given in Fig. \ref{fig:tsky} from the convolution of the ideal array factor pattern with a model of the sky emission.

The combination of the large absorber, the directivity of the relatively large array, and limiting measurements to when the Galctic centre was below the horizon leads to an estimate of less than 1~K for the stray contribution $T_{\text{ext},\text{sky}(B)}$ from the sky beyond the edges of the absorber.  The Sun was between 31$^\circ$ and 44$^\circ$ elevation for all measurements.  It did not contribute significantly to $T_{\text{ext},\text{sky}(A)}$ or uncertainty therein as it was not high enough to enter the main beam or near sidelobes.

The ripple in Fig. \ref{fig:result} has an amplitude of 3~K to 4~K and a period of approximately 100~MHz that is consistent with a standing wave associated with the 3~m~$\times$~3~m absorber load. Plots of beamformed spectra show that, below 1.2~GHz, the ripple is seen only in the absorber-load spectrum and not in the sky spectrum. This suggests the ripple is introduced by the load and not the PAF. Above 1.2~GHz the load and sky spectra are both more complicated and it is harder to say where the ripple comes from without further experiments.

Gaps in the frequency coverage of Fig. \ref{fig:result} are due to intermittent faults in the calculation of the array covariance matrices by the beamformer.  Matrices that were not Hermitian or not positive-semidefinite were discarded.  We have since fixed the beamformer firmware to remove this problem.  Two malfunctioning array ports were excluded from analysis by removing corresponding rows and columns from the covariance matrices before further analysis. One large negative spike was removed from the final noise temperature curves, as it was likely due to narrow-band radio-frequency interference. 

\begin{table*}
\caption{Measurement parameters.}
\label{tab:measparams}
\begin{center}
\footnotesize
\begin{tabular}{ccccccc}
\toprule
  Band & Sampled & Digtal Centre         & Beamformer  & Observation & LST & Absorber \\ 
       & RF Band & Frequency & Band & UTC Epoch & Range & Temperature \\
       &  (MHz)          & (MHz) & (MHz)& (yyyymmddhhmmss) & (hh:mm) & ($^\circ$C) \\ \midrule
  1 & 700-1200 & 891 &   700-1083 & 20140728013215 & 05:41-06:10  & 17$\pm2$ \\
  2 & 840-1440 & 1243 & 1052-1435 & 20140728024948 & 06:59-07:27  & 21$\pm2$ \\
  3 & 1400-1800 & 1603 & 1412-1795 & 20140728040738 & 08:17-08:44  & 24$\pm2$\\
  \bottomrule
\end{tabular}
\end{center}
\end{table*}

\section{Conclusion}
Our aperture-array noise temperature measurements for the first Mk.~II ASKAP PAF show a doubling of the bandwidth over which low noise temperature is achieved compared to the Mk. I PAF results in \cite{Schinckel2011}.  Above 1.4~GHz, sensitivity as an aperture array has been doubled.  This same improvement was seen in on-dish testing \cite{Chippendale2015}, so astronomical surveys above 1.4~GHz should be four times faster when made with Mk.~II PAFs replacing equivalent numbers of Mk. I PAFs.

Aside from the 4~K amplitude ripple, our analysis suggests that measurement uncertainty from other known sources is less than 1.2~K.  This improvement on our earlier achievement of 4~K uncertainty \cite{Chippendale2014} comes from using a larger absorber box closer to a larger array.  However, we need to understand and mitigate the ripple in Fig. \ref{fig:result} before we can truly claim 1.2~K accuracy.  We have commenced work towards this by measuring the absorber box at three different heights over the array under test.  Lining the inside and lower edges of the box's side walls with absorber may also help.


\section*{Acknowledgment}
The Australian SKA Pathfinder is part of the Australia Telescope National Facility which is managed by CSIRO. Operation of ASKAP is funded by the Australian Government with support from the National Collaborative Research Infrastructure Strategy. Establishment of the Murchison Radio-astronomy Observatory was funded by the Australian Government and the Government of Western Australia.  ASKAP uses advanced supercomputing resources at the Pawsey Supercomputing Centre. We acknowledge the Wajarri Yamatji people as the traditional owners of the Observatory site. 

Creating the Mk.~II ASKAP PAF was a team effort involving many beyond the authors.  This included A. Schinckel as ASKAP Project Director and Dr. A. Rispler as Project Manager.  Dr. J. Bunton provided key concepts and specifications across the system as ASKAP Project Engineer.   The  prototype team included S. Broadhurst, W. Chandler, D. Chandler, P. Doherty, D. Kiraly, J. Kanapathippillai, N. Morison and Dr. P. Roberts.  The firmware team included S. Neuhold (manager), Dr. J. Tuthill, T. Bateman, C. Haskins and Dr. J. Bunton.  The production team included S. Barker (leader), M. Shields (project engineer), Alan Ng, Andrew Ng, A. Sanders, R. Chekkala, D. Kiraly, W. Cheng and N. Morison.  Electromagnetic design and validation was supported by R. Gough.  The CSIRO Marsfield Workshop supported the prototype work, in particular P. Cooper, M. Bourne, M. Hyunh, R. Moncay and M. Death.  Additional assistance was given to the prototype work by S. Castillo, Y. Chung, D. Gain, L. Li, S. Mackay and L. Reilly.  Deployment and testing at the MRO were supported by S. Jackson, M. Reay, J. Morris, L. Puls, R. McConigley, Dr. S. Amy and B. Hiscock. 




\bibliographystyle{IEEEtran}
\bibliography{IEEEabrv,apj-jour,ade_aperture_array_sens_isap_2015}
%
%
%

\end{document}